\documentclass[aps,floatfix,superscriptaddress,notitlepage,nofootinbib]{revtex4-1}
\usepackage{amsmath,amssymb,amsfonts,graphics,graphicx,dcolumn,bm,enumerate}
\usepackage{comment,natbib,appendix}
\usepackage{multirow,color}
\usepackage{chngpage}
\usepackage{afterpage}
\usepackage{xcolor}
\usepackage{amsthm}
\usepackage{natbib}
\usepackage{hyperref}
\usepackage[margin=1.2in]{geometry}
\usepackage{epstopdf}
\usepackage{float}

\newcommand{\cmp}
{\affiliation{Saha Institute of Nuclear Physics, Kolkata 700064, India.}}
\newcommand{\isi}
{\affiliation{Economic Research Unit, Indian Statistical Institute, Kolkata 700108, India.}}
\newcommand{\raghunathpur}
{\affiliation{Department of Physics, Raghunathpur College, Raghunathpur, Purulia - 723133, West Bengal, India.}}
\newcommand{\SRM}
{\affiliation{Department of Physics, SRM University-AP, Andhra Pradesh - 522240, India}}

\begin{document}
\title{Scaling Behavior of the Hirsch Index for Failure Avalanches, Percolation
Clusters and Paper Citations}

\author{Asim Ghosh}
\email[Email: ]{asimghosh066@gmail.com}
\raghunathpur
 
 \author{Bikas K. Chakrabarti }%
 \email[Email: ]{bikask.chakrabarti@saha.ac.in}
 \cmp \isi 
 
\author{Dachepalli R. S. Ram}
\email[Email: ]{dachepalli\_ravi@srmap.edu.in}
\SRM
 
\author{Manipushpak Mitra}
\email[email: ]{manipushpak.mitra@gmail.com}
\thanks{corresponding author}
\isi

\author{Raju Maiti }
\email[Email: ]{rajumaiti@gmail.com}
\isi

\author{Soumyajyoti  Biswas}
\email[Email: ]{soumyajyoti.b@srmap.edu.in}
\SRM

\author{Suchismita Banerjee}
\email[Email: ]{suchib.1993@gmail.com}
\isi

\begin{abstract}
A popular measure for 
citation inequalities of individual
scientists has been the Hirsch index
($h$).
If for any scientist the number $n_c$ of citations is plotted against the serial number $n_p$ of the paper having those many citations (when the papers are ordered from highest cited to lowest) then $h$ corresponds to the nearest lower integer value of $n_p$ below the fixed point of the non-linear citation function (or given by $n_c = h = n_p$ if both $n_p$ and $n_c$ are dense set of integers near the $h$ value). 
The same index can be estimated (from $h=s=n_{s}$) for the avalanche or cluster of size ($s$) distributions ($n_s$) in elastic fiber bundle or percolation models.
Another such inequality index, called the Kolkata index ($k$) says that $(1-k)$ fraction of papers attract $k$ fraction of citations ($k=0.80$ corresponds to the 80-20 law of Pareto).
We find, for stress  ($\sigma$), lattice occupation probability ($p$) or Kolkata index ($k$) near the bundle failure threshold ($\sigma_c$) or percolation threshold ($p_c$) or critical value of Kolkata index $k_c$,  good fit to Widom-Stauffer like scaling $h/[\sqrt{N}/log N]$ = $f(\sqrt{N}[\sigma_c -\sigma]^\alpha)$, $h/[\sqrt{N}/log N]=f(\sqrt{N}|p_c -p|^\alpha)$ or $h/[\sqrt{N_c}/log N_c]=f(\sqrt{N_c}|k_c -k|^\alpha)$ respectively, with asymptotically defined scaling function $f$, for systems of size $N$ (total number of fibers or lattice sites)  or $N_c$ (total number of citations), and $\alpha$ denoting the appropriate  scaling exponent.  We also show that if  the number ($N_m$) of  members of parliaments or national assemblies of
different countries (with population $N$) is identified as  their respective $h-$index, then the data fit the scaling relation $N_m \sim \sqrt N /log N$, resolving a major recent controversy.
\end{abstract}
 
 \maketitle
 
\section{Introduction}
Monotonically and non linearly decaying inequality functions are ubiquitous. When the number ($n_p$) of papers by any author (or,for that matter, an institution) are arranged according to the number ($n_c$) of the citations they received, the citation inequality function becomes a monotonically decaying one (see e.g., \cite{hirsch2005index}).
The same is true  for avalanches in materials failure or in earthquakes (see e.g., \cite{biswas2015statistical}), cluster size distributions in percolation problems (see e.g., \cite{stauffer2018introduction}), etc., where the number $(n_s)$  of avalanches or clusters (giving the size inequality function) decrease monotonically and nonlinearly with the size ($s$) of the avalanche or cluster.
Large avalanches, strong quakes, or big size clusters come or occur in small numbers, while the smaller or weaker the avalanches or quakes, or clusters, the larger are their abundance in occurrence.
How does one statistically measure these inequalities in  occurrence frequencies of citation numbers or avalanche or cluster sizes? Obviously, the corresponding distribution functions for inequalities in citations or sizes would contain the entire statistics.
However, they are not convenient to handle. One can consider the citation number of the best cited paper (as sometimes done for some unique awards, etc.), or study the statistical (self-similar) structure of the biggest avalanche or the largest (percolating) cluster (as in statistical physics \cite{stauffer2018introduction}). 
Hirsch proposed  \cite{hirsch2005index} an index to measure these inequalities, by locating the fixed point of the nonlinear inequality function. 
The Hirsch index ($h$) corresponds to the citation number or occurrence frequency which is  commensurate in magnitude with the number of publications or avalanche cluster sizes.

Systems near their critical points, self-organized or tuned, have mostly been studied in the self-similar limit of their divergent correlations (see e.g., \cite{biswas2015statistical, stauffer2018introduction}).
The corresponding critical exponent values arising out of such self-similarities have helped to classify vastly different physical systems based on their symmetries, dimensionality and such broad qualifiers. 
One powerful tool has been the scaling relations among the critical exponents that helped to build an
interconnected and precise relation between experimental observables in such systems near criticality. 
In this work, however, we focus on quantifying the response of near-critical systems through the corresponding inequality statistics (for example, in  $n_c$ versus $n_p$ of citations or in $n_s$ versus $s$ for avalanches or clusters). Specifically, we measure the widely used Hirsch index ($h$), which in effect, gives a measure of a size that is commensurate with its relative abundance.
It turns out to be possibly even more robust than the critical behavior (characterized by a set of exponents). We demonstrate this by choosing a wide variety of systems: citation statistics, percolation cluster statistics and avalanche statistics in Fiber Bundle Models (FBMs) and even in the statistics of Parliament sizes in different countries of the world. 
They differ in their dimensionality (two dimensions for the percolation model studied here, mean-field for the fracture model, possibly small-world networks for citation and parliament statistics). They further differ in their measurement variables, and their size distribution statistics, making them widely different in terms of their prominently apparent features. 
However, we show here that in spite of their obvious differences, the scaling behavior of the Hirsch index shows remarkable universality.


\begin{figure}[H]
\centering
\includegraphics[width=0.7\textwidth]{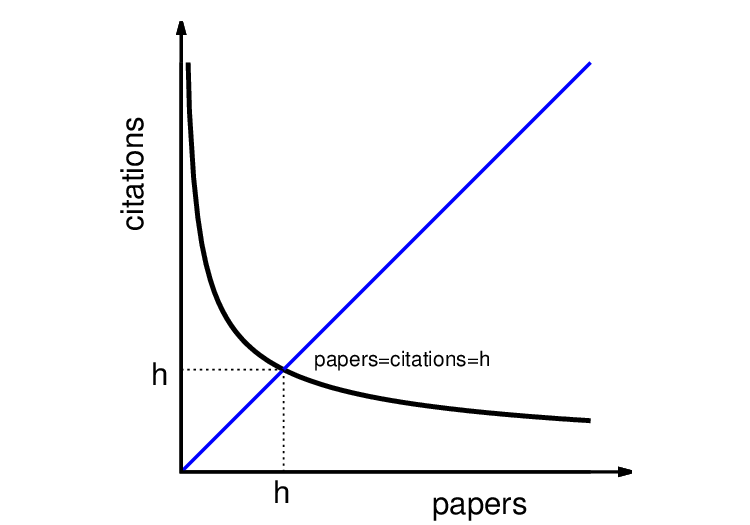}
\caption{A schematic drawing of the citation function of a typical scientist.
$h$-index (an integer) is given by lower value of the paper serial number
below the fixed point value (intersection point of the 45 degree line
from the origin), when the papers are ordered from highest cited to the
lowest cited one. When the citation number and the serial number of the paper
(having those citations) are both sequential integers near the fixed
point, the citation number equals the number of papers and both become
$h$. Similar will be the case where citations are replaced by the failure of the materials avalanche sizes (or cluster sizes) and the paper numbers
are replaced by the number of such avalanches (or of the clusters in pre- or post-percolating systems).} 
\label{fig1}
\end{figure}


How does the $h$ index scale with a total number of
publications ($N_p$) by the author (institution) or
the total number ($N_c$) of citations received by
the author (institution)? Young \cite{yong2014critique} suggested
analytically that $h$-index value should scale
with the total number of citations $N_c$ ($=\sum n_c$) as
$\sqrt N_c$ asymptotically. In ref. \cite{ghosh2021limiting}, a brief
analysis of the Google Scholar data indicated
independently that $h \sim \sqrt N_p$ for large values of $N_p$ which corresponds to the highest value of
$n_p$, the total number of publications by the author.  Note that if
both these relationships are valid then,
statistically speaking, for a prolific author,
the total number of citations would be linearly
proportional to the total number of papers published
and the proportionality constant would perhaps be
determined by the size of the existing author
network in the subject (suggesting an effective
Dunbar number \cite{dunbar1992neocortex,bhattacharya2016sex}, for the community of authors).

In a recent Monte Carlo study \cite{biswas2021social} on the avalanche
sizes and their numbers in the Fiber Bundle Models (FBM) of materials failure (see e.g.,
\cite{biswas2015statistical,pradhan2010failure}) due to increasing stress 
on such bundles, the numerical analysis of the data
for the nonlinearly decaying numbers of avalanches
with their sizes (or released elastic energies)
suggested $h \sim \sqrt N/{log N}$.

Our numerical study here of avalanche size distributions
in FBMs (for stress or load per fiber $\sigma$ less than
its global failure stress $\sigma_c$), of cluster size distributions for lattice occupation concentration $p$ near
the percolation point $p_c$ (for $p$ both below and above
$p_c$) in the percolating system and the previous analysis \cite{ghosh2021limiting} of citation distributions of scientists
with Kolkata index $k$ (see e.g, \cite{banerjee2020inequality} for a review, giving the fraction $k$
of citations/wealth attracted/possessed by $1 - k$ fraction of
publications /people) near (both above and below) the threshold point $k_c
(= 0.86)$, all show excellent fit to a Widom-Stauffer like scaling relation between the Hirsch index ($h$)
and system size $N$ (or individual’s total citation size $N_c$), following
Widom Scaling for the free energy away from the critical point and the
subsequent Stauffer Scaling \cite{stauffer2018introduction} for the number of a particular sized
cluster, identified here as the equivalent Hirsch index, at and away from the percolation
point), 
\begin{subequations}
\begin{align}
\frac{h}{[\sqrt N/log N]} & = f(\sqrt{N}[\sigma_c -\sigma]^\alpha) \\
\frac{h}{[\sqrt N/log N]} & = f(\sqrt{N}|p_c - p|^\alpha)\\
\frac{h}{[\sqrt N/log N]} & = f({\sqrt N_c} |k_c - k|^{\alpha}),
\end{align}
\end{subequations}
\noindent with asymptotically well-defined finite size scaling function $f$
($f(x)$ = constant at $x = 0$ and $f$ remaining continuous and  finite as $x$ approaches infinity) for systems of size $N$ (total number of
fibers or lattice sites/bonds) or size $N_c$ (total number of citations of
all the publications by an individual scientist), and $\alpha$ denoting
the appropriate scaling exponent.

Traditionally the Hirsch index $h$ for different authors have been  fitted [4] to the scaling form $\sqrt N_c$. For cases where the Kolkata index values of the authors are not known, we simply fitted, following the scaling relation (1) to the form appropriate for the critical point
\begin{subequations}
\begin{align}
h &\sim  \sqrt N_c /log N_c \\
  &\sim \sqrt N_p /log N_p.
\end{align}
\end{subequations}
\noindent We also show  that if  the
number ($N_m$) of  members  of the
parliaments or national assemblies of
different countries (with the corresponding population denoted by  $N$) are identified as  those countries' respective $h$
indices, then the data fit well to the scaling relation $N_m \sim \sqrt N /log N$.
This helps to comprehend the discrepancies (c.f. \cite{taagepera1972size}) with the $\sqrt N$ relationship, reported in a very recent analysis 
\cite{margaritondo2021size}, resolving a recent major controversy.

\section{Data Analysis \& Numerical Studies}
\label{sec2}
\subsection{Failure avalanches in fiber bundle model (FBM)}
The fiber bundle model (FBM) is a generic model for failure of disordered solids. 
An ensemble of $N$ fibers are set between two rigid parallel plates and a load is
applied on the bottom plate. Each fiber is linear elastic, with the same elastic constant
and has a failure threshold selected randomly from a distribution. This failure threshold
is the source of disorder and non-linearity in the otherwise linear model.

\begin{figure}[H]
\centering
\includegraphics[width=0.7\textwidth]{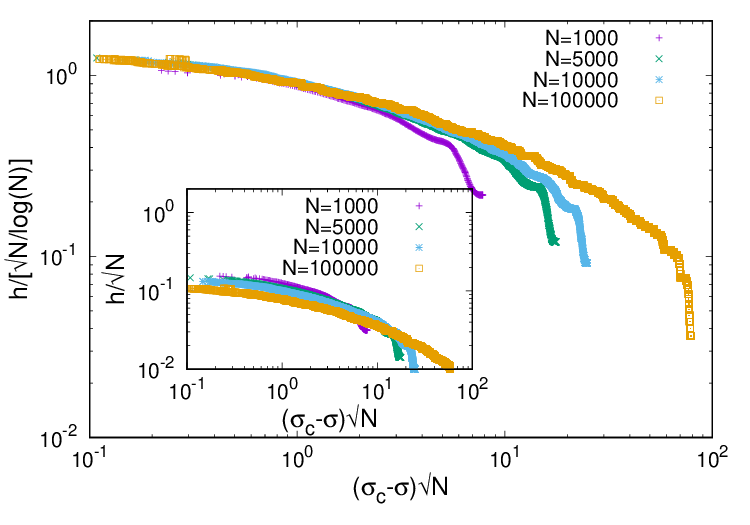}
\caption{Finite size scaling analysis of the fiber bundle model $h-$index (from avalanche size distribution). The scaling fit to Widom-Stauffer relations (1) are obtained by making  all the data points of different sizes ($N=1000, 5000, 10000$ and $100000$) and stress level (different $\sigma$ values)  collapse together.  It turns out that the scaling form (1b) with exponent $\alpha=1$ gives good data collapse. The inset shows that the  data collapse clearly gets worsened by dropping the $log N$ term in the scaling relation (1b).}
\label{fig2}
\end{figure}

When a small load ($W$) is applied, the weakest fiber breaks, and the load carried by that
fiber is shared equally by all the remaining fibers, which can trigger further
failures. Through gradual increase of load, therefore, the model goes through intermittent
stable states, which are subsequently perturbed by increasing the load slowly. In going from
one stable state to another, the number of fibers that break is the avalanche size ($s$).
The size distribution of this follows a power-law statistics $P(s)\sim s^{-5/2}$ for $s\longrightarrow\infty$   \cite{hansen1992distribution}. The
avalanche dynamics continues until the load per fiber value $\sigma=W/N$ reaches a critical
limit $\sigma_c$, when the entire system breaks down. 

The avalanches are arranged in the ascending order to estimate $h$ index values for stress level $\sigma$ below $\sigma_c$. It was shown (\cite{biswas2021social}) that the terminal value of
$h$ ($=h_f$)  at the critical point ($\sigma=\sigma_c$) follows a scaling relation $h_f\sim \sqrt{N}/log(N)$. Here
we look for the scaling of $h$ with the critical interval ($\sigma_c-\sigma$). It turns out that the
scaling relation (1a) fits very well with our numerical the data 
(see Fig. \ref{fig2}). We also attempted the scaling fit without the $log(N)$ term in the above mentioned relation (see inset of Fig. \ref{fig2}) and it clearly  worsens the  data collapse.

\begin{figure}[H]
\centering
\includegraphics[width=0.7\textwidth]{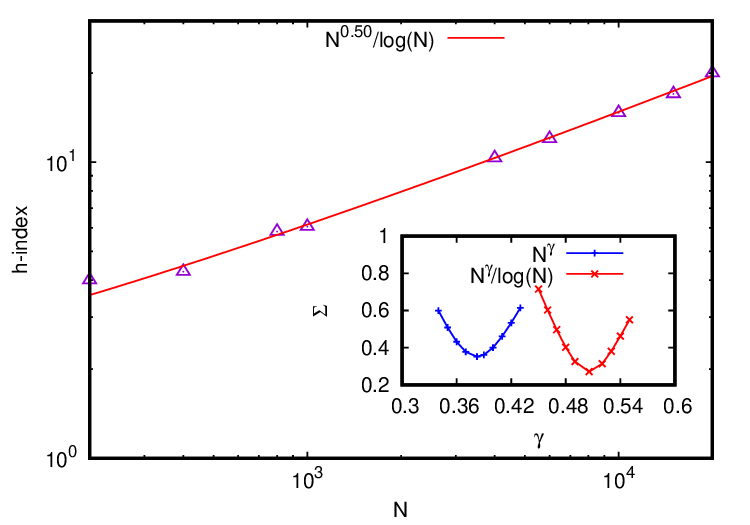}
\caption{Scaling behavior for the average $h$-index
(in the range $4 \le h \le 20 $) at the
breaking point ($\sigma=\sigma_c$) of the bundles with total
number $N$ (in the range $200 \le N \le 20000 $) of
 fibers in the equal-load-sharing FBM (with
uniform distribution of fiber breaking thresholds)
considered here (cf. \cite{biswas2021social}).
The Fig. shows the  best fit of $h$ to
$\sqrt{N}/log N$. The inset shows statistical deviations $\Sigma = \sqrt{\Sigma_i (h_d(i) - h_\gamma(i))^2/\Sigma_i 1}$ to the scaling forms  $h\sim N^{\gamma}$ and  $h\sim N^{\gamma}/log (N)$ with respect to $\gamma$, where
$i$ represents the individual (random) realization of the  FBM  with the
corresponding $h$  ($=h_d$) value obtained from the  simulation result and  the scaling fit ($h_\gamma$) from the above mentioned scaling relations. The inset shows
that the statistical deviation  becomes minimum for the scaling
with log correction term with the appropriate value of $\gamma$
near 1/2.}
\label{fig3}
\end{figure}

The Fig.\ref{fig3} shows the scaling behavior for the average $h$-index
(in the range $4 \le h \le 20 $) at the
breaking point ($\sigma=\sigma_c$) of the bundles with total
number $N$ (in the range $200 \le N \le 20000 $) of
fibers in the equal-load-sharing FBM (with
uniform distribution of fiber breaking thresholds)
considered here (cf. \cite{biswas2021social}).
The Fig. shows the  best fit of $h$ to
$\sqrt{N}/log N$.

\begin{figure}[H]
\centering
\includegraphics[width=0.7\textwidth]{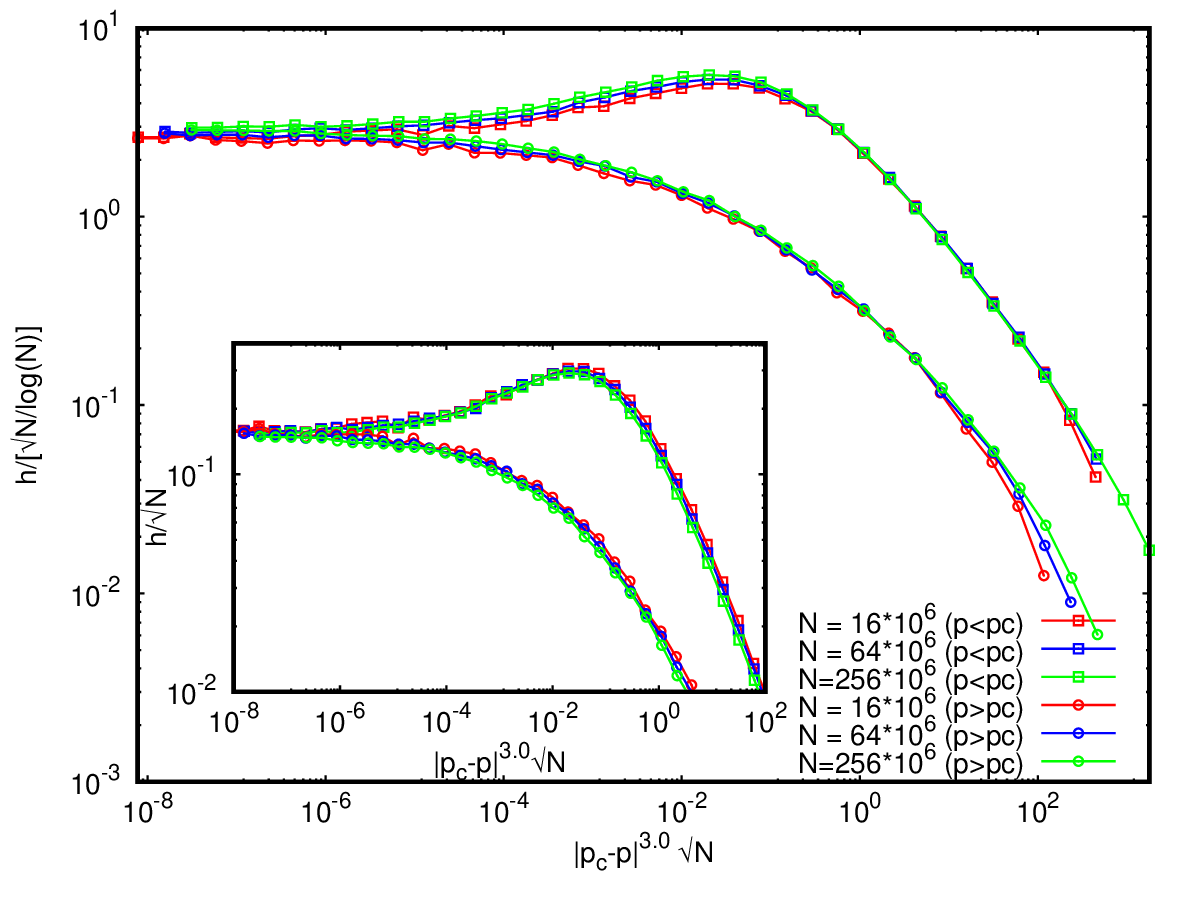}
\caption{Finite size scaling analysis of the 2D site percolation $h-$index (from the cluster size distribution). The scaling fit to Widom-Stauffer relations (1) are obtained by making  all the data points of different sizes ($N=4000\times4000, 8000\times8000$ and $16000\times16000$) and concentrations (different $p$ values, $p_c=0.5927$)  collapse together.  It turns out that the scaling form (1b) with exponent $\alpha=3$ gives good data collapse. The inset shows that the  data collapse seems to get worsened by dropping the $log N$ term in the scaling relation (1b).}
\label{fig4}
\end{figure}

\subsection{Percolation clusters in 2D lattice}
Here we consider two-dimensional (2D) site percolation in a square lattice with site occupancy probability $p$ ($0<p<1$). For a given $p$, we measure  cluster distribution, and hence the h-index is computed (here h-index measures $h$ number of clusters each having cluster sizes greater than equal to $h$).   In our simulation, we took four different system sizes ($N=4000\times4000, 8000\times8000$ and $16000\times16000$) and h-index was estimated for different $p$ values. 

Here we carry out the finite size scaling analysis \cite{stauffer1979scaling} (see also \cite{manna2022near}) of the $h$ index for different system sizes. The critical exponent $\alpha$ is determined by fit to scaling relation (1b) for  all the data points  of different sizes ($N=4000\times4000, 8000\times8000$ and $16000\times16000$). We obtain good data collapse with $\alpha=3$ and different scaling function $f$  for $p>p_c$ and $p<p_c$ while they converge to the same value at $p=p_c$ (see Fig \ref{fig4}).    The inset of Fig. \ref{fig4} shows the  scaling fit to relation (1b) without the $log(N)$ term and it is observed that right hand side scaling fit is not that well.

\begin{figure}[H]
\centering
\includegraphics[width=0.7\textwidth]{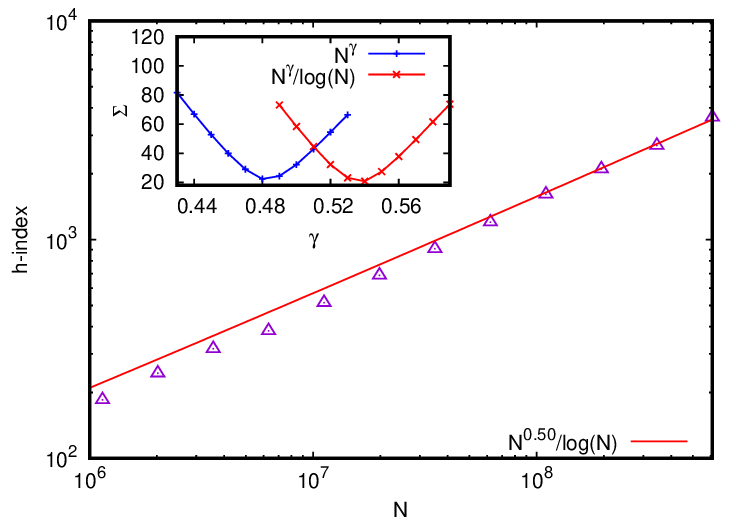}
\caption{Scaling behavior for the $h$-index (in the range $185 \le h \le 4421 $) for cluster size distributions at the percolation point  (with  size $N$ on square lattice; $10^6<N<10^9$).
The Fig. shows the best fit of $h$ to
$\sqrt{N} /log N$. The inset shows the statistical deviations $\Sigma = \sqrt{\Sigma_i (h_d(i) - h_\gamma(i))^2/\Sigma_i 1}$ to the scaling forms  $h\sim N^{\gamma}$ and  $h\sim N^{\gamma}/log (N)$ with respect to $\gamma$, where
$i$ represents the individual (random) realization of the  percolating system  with the
corresponding $h$  ($=h_d$) value obtained from the  simulation result and  the scaling fit ($h_\gamma$) from the above mentioned scaling relations. The inset shows
that the statistical deviation  becomes minimum for the scaling
with log correction term, though with a value of $\gamma$ somewhat above 1/2.}
\label{fig5}
\end{figure}

The cluster size distribution
in the percolation problem has also been studied
for estimating the $h$-index scaling with the total
number $N$ of lattice sites at the percolation
threshold of site percolation on square lattice.
We of course find here the best fit scaling form
to be $h \sim \sqrt{N}/log N$ (see Fig. \ref{fig5}).

\subsection{Paper citations}
We first analyze the data for $h$-index and its
scaling with the total number of publications $N_p$
and of citations $N_c$ for the one hundred
scientists (in mathematics, physics, chemistry,
medicine, biology, economics, sociology; including
those of twenty Nobel Laureates in those subjects),
given in ref. \cite{ghosh2021limiting}. Next, we collected (in May-June,
2021) the same kind of data\footnote{The data will be available on request to corresponding author} for one thousand
scientists  (mostly physicists) in all the above-mentioned subjects
from Google Scholar. 

\begin{figure}[H]
\centering
\includegraphics[width=0.6\textwidth]{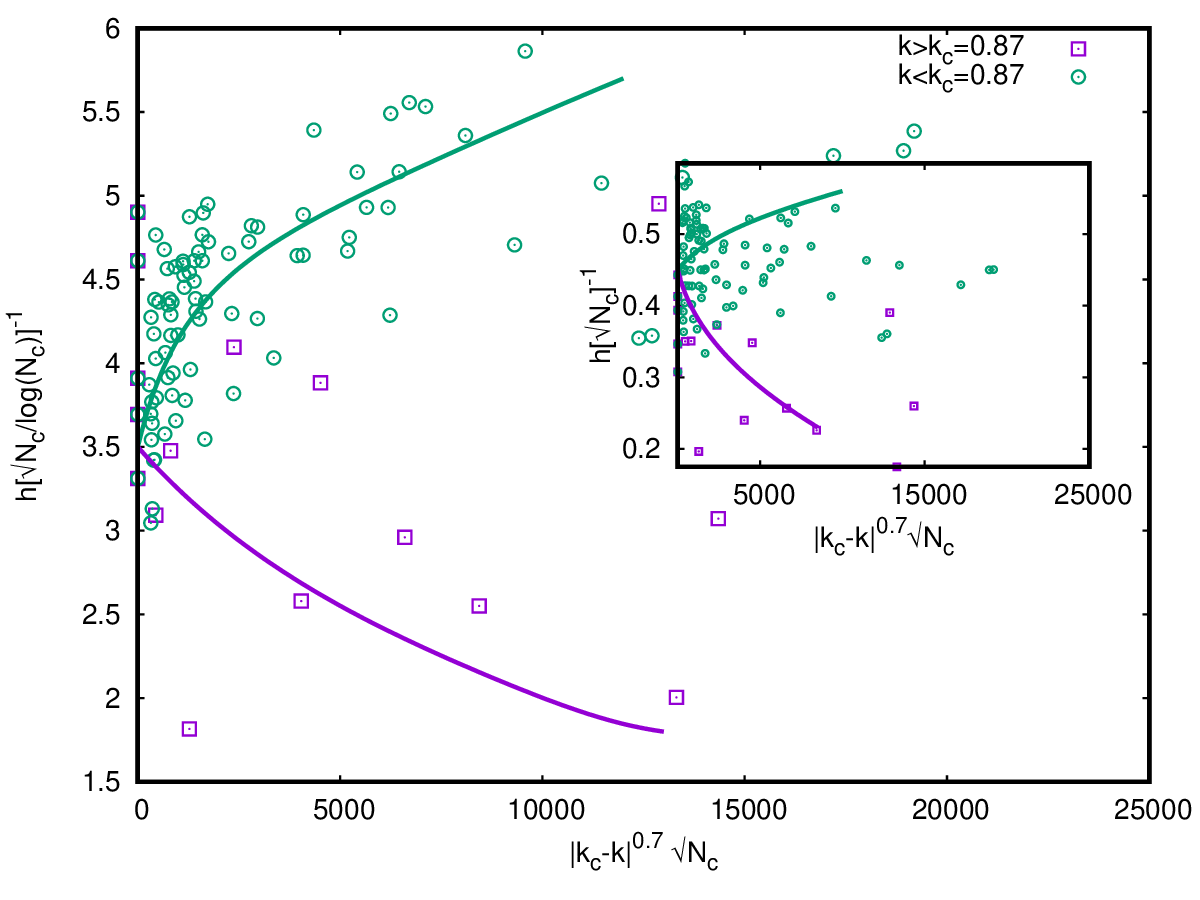}
\caption{Finite size ($N_c$)  scaling analysis of the  citation
distributions of 100 individual scientists (table 1 of Ref. \cite{ghosh2021limiting}).
The scaling fit to Widom-Stauffer relation (1c) is obtained by
fitting all the data for their $h$ index and the $N_c$ values  of total citations
of the publications by individual scientists and their corresponding
Kolkata index values $k$ near the critical value $k_c \simeq 0.87$
\cite{ghosh2021limiting}. It turns out that the scaling form (1c) with
exponent $\alpha=0.7$ gives good data collapse. The inset shows that
the data collapse seems to get worsened by dropping the $logN$ term
in the scaling relation (1c), though at the critical  point $k = k_c$,
one gets $h = a  \sqrt {N_c}$, with $a = 0.45
\pm 0.10$, which includes the analytical prediction \cite{yong2014critique} that
$a = 0.54$ in the large $N_c$ limit. The continuous lines (in both  the main figure and the inset) represent the scaling functions for $k<k_c$ (green) and $k>k_c$ (violet), and  are drawn for a guide to the eye.}
\label{fig6new}
\end{figure}

\begin{figure}[H]
\centering
\includegraphics[width=0.6\textwidth]{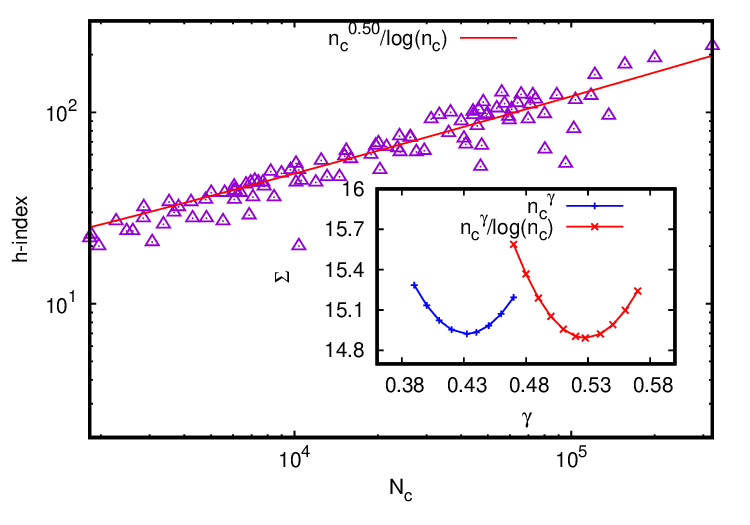}
\caption{Scaling behavior for the $h$-index (in the range
$20 \le h \le 222 $) of the 100 scientists 
(data  taken from \cite{ghosh2021limiting}), with the total number of
citations $N_c$ (in the range $1819 \le N_c \le 323473$). Our analysis shows the best fit to
$h \sim \sqrt{N_c} /log N_c$. The inset shows the statistical deviations $\Sigma = \sqrt{\Sigma_i (h_d(i) - h_\gamma(i))^2/\Sigma_i 1}$ to the scaling forms  $h\sim N_c^{\gamma}$ and  $h\sim N_c^{\gamma}/log (N_c)$ with respect to $\gamma$, where
$i$ represents the individual scientist  with the
corresponding $h$  ($=h_d$)  obtained from Google Scholar and  the scaling fit ($h_\gamma$) obtained from the above mentioned scaling relations. The inset shows that the
statistical deviation  becomes minimum for the scaling
with log correction term with the value of $\gamma$
a little higher than 1/2.}
\label{fig6}
\end{figure}

Figs. \ref{fig6new} and \ref{fig6} show the scaling behavior for the $h$-index (in the range
$20 \le h \le 222 $) of the 100 scientists 
(data  taken from \cite{ghosh2021limiting}), with the total number of
citations $N_c$ (in the range $1819 \le N_c \le 323473$). Our analysis shows the best fit to
$h \sim \sqrt{N_c} /log N_c$.

\begin{figure}[H]
\centering
\includegraphics[width=0.7\textwidth]{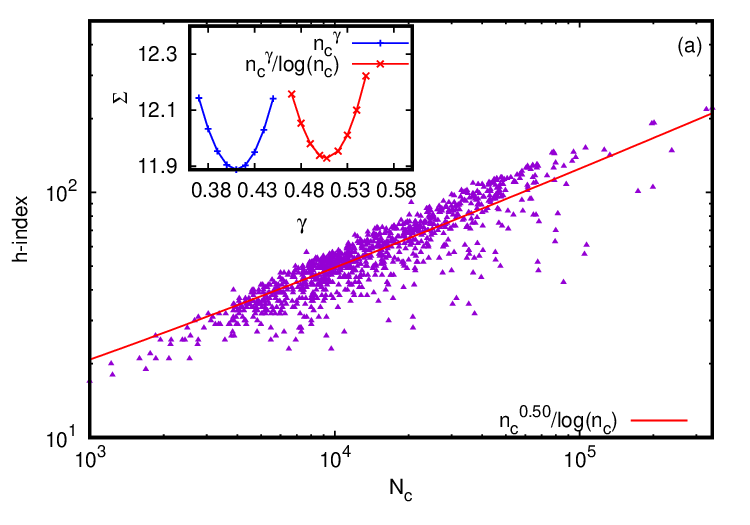}
\includegraphics[width=0.7\textwidth]{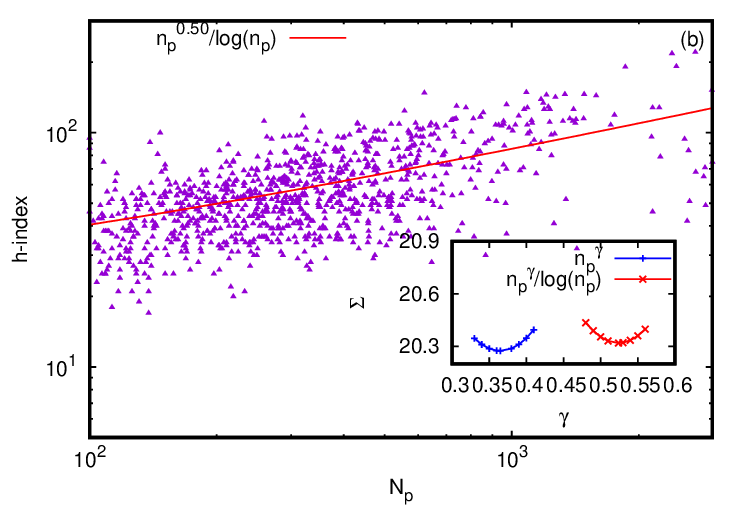}
\caption{Scaling behavior for the $h$-index (in the range
$ 17\le h \le 221$) of the 1000 scientists with the total number of
citations $N_c$ (in the range $996\le N_c \le 348680$)
in (a) and with the total number of papers $N_p$
(in the range $100 \le N_p \le 2987$) in (b). The data are taken from  Google
Scholar in June 2021. The Figs. show the best
fits to $h \sim \sqrt N_c /log N_c$ in (a) and to
$h \sim \sqrt N_p /log N_p$ in (b). The insets show the statistical deviations $\Sigma = \sqrt{\Sigma_i (h_d(i) - h_\gamma(i))^2/\Sigma_i 1}$ to the scaling forms $h\sim N_p^{\gamma}$ (or $h\sim N_c^{\gamma}$) and  $h\sim N_p^{\gamma}/log (N_p)$ (or $h\sim N_c^{\gamma}/log (N_c)$) with respect to $\gamma$, where
$i$ represents the individual scientist  with the
corresponding $h$  ($=h_d$) value obtained from Google Scholar and  the scaling fit ($h_\gamma$) from the above mentioned scaling relations. The inset
shows that though the statistical deviation  does not become
minimum for the scaling with log correction term (presumably
due to inclusion of some tail-end points), the best fit for
$\gamma$ assumes the desired value 1/2.}
\label{fig7}
\end{figure}

The Figs \ref{fig7}(a,b) scaling behavior for the $h$-index (in the range
$ 17\le h \le 221$) of the 1000 scientists with the total number of
citations $N_c$ (in the range $996\le N_c \le 348680$)
in (a) and with the total number of papers $N_p$
(in the range $100 \le N_p \le 2987$) in (b). The data are taken from  Google
Scholar in June 2021. The Figs. show the best
fits to $h \sim \sqrt N_c /log N_c$ in (a) and to
$h \sim \sqrt N_p /log N_p$ in (b). 

\subsection{Number of the representatives in the National Assemblies}
\begin{figure}[H]
\centering
\includegraphics[width=0.7\textwidth]{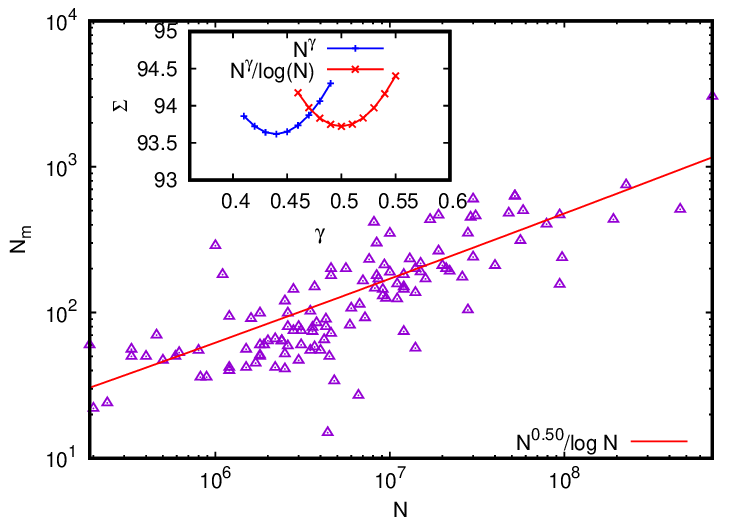}
\caption{Scaling behavior for the number $N_m$ (in
the range $15 \le N_m \le 3040$)  of the
representatives in National assemblies or
Parliaments  of different countries of the
world (1972 data from the original study
\cite{taagepera1972size}; see also \cite{margaritondo2021size})    with total population $N$ (in the range
$19\times10^4 \le N \le 70.5\times10^7$) of the respective countries. When $N_m$
is identified as the $h-$index for the
countries, we get
$N_m\sim \sqrt N /log N_m$ which is fitted here with the data. The inset shows  the statistical deviations $\Sigma = \sqrt{\Sigma_i (N_{md}(i) - N_{m\gamma}(i))^2/\Sigma_i 1}$ to the scaling forms  $N_m\sim N^{\gamma}$ and  $N_m\sim N^{\gamma}/log (N)$ with respect to $\gamma$, where
$i$ represents the individual country  with the
corresponding members
of parliaments $N_m$  ($=N_{md}$)  and  the scaling fit ($N_{m\gamma}$) from the above mentioned scaling relations.
This resolves the discrepancy noted in the
analysis of the same data in ref. \cite{margaritondo2021size}. Here again, as
seen from the inset, though the statistical deviation  does not
become minimum for the scaling with log correction term
(presumably due to inclusion of some tail-end points), the
best fit for $\gamma$ assumes the desired value 1/2.}
\label{fig8}
\end{figure}

Finally, we note that if the number $N_m$  of the
representatives in the National Assemblies or in
the Parliaments  of different countries of
the world are identified  as the $h$-index for
respective country, having population $N$, then
we find (see Fig. \ref{fig8}) that the data analyzed in
ref \cite{taagepera1972size}\footnote{The data extracted from
this paper can be found in
https://sciencehistory.epfl.ch/physics-and-sociology/}
shows the best fit of $N_m$ to $\sqrt N/log N$. This
also resolves the discrepancy noted  in the analysis if the same data (\cite{margaritondo2021size} and references therein).

\section{Summary and Conclusion}
We have studied here the scaling behaviors of Hirsch index $h$ 
for inequalities \citep{hirsch2005index}  applied to
the unequal distributions of responses and statistics in different physical (fracture and percolation) and social systems (citations and parliament sizes). 
We have shown that Hirsch index follow remarkable off-critical Widom-Stauffer scaling in these sytems that are
widely different in terms of their dimensionality and symmetry.

We have studied avalanche or
cluster sizes in physical systems like the fiber
bundle models (see e.g., \citep{biswas2015statistical}\citep{pradhan2010failure}) and percolating
systems (see e.g., \cite{stauffer2018introduction,stauffer1979scaling}). Indeed, surprising
successes  of such Hirsch-like social inequality
measures were already seen (see e.g., \citep{biswas2021social,manna2022near}) in predicting  the global failures in fiber
bundles and the self-organized critical points of sand-pile systems.
We show here, from the Monte Carlo simulation
results (see sections IIA and IIB), that the
Widom-Stauffer-like  scaling relation (1a,b) of the
Hirsch index $h$ (defined here for the size
$s$ distributions $n_s$, through $h = s = n_s$)
fit remarkably well with the system size $N$
(see Figs. 2 and 4). We also show (see the insets
of Figs. 2 and 4) that the scaling collapse breaks
down without the $log N$ term in (1) (visibly clear
for the fiber bundle results in Fig. 2).

As is well known, the Hirsch index ($h$) was introduced originally 
to measure  the
inequalities in success (through citations in the
subsequent literature) of the contributions
(papers) of individual scientists. The citation
function (see Fig. 1), is an well documented
nonlinear and monotonically decaying function
(cf. Zipf law \cite{zipf1949human}). $h$
corresponds to the fixed point  of this
 citation function. Analytical study
 \cite{yong2014critique}, as well as numerical
data analysis (see e.g., \cite{ghosh2021limiting}),
 suggested $h \sim \sqrt {N_c}
\sim \sqrt {N_p}$, for any author having
$N_p$ total papers and
$N_c$ total citations. Since
for different scientists, we do not have
any knowledge of the critical intervals (like
$|\sigma_c - \sigma|$ or $|p_c-p|$ in relations (1)) here the  detailed data
for $h$ are fitted to the relations (2)  (see section IIC,
Figs. \ref{fig6}, \ref{fig7}(a,b)) assuming
the critical interval to be zero for the
chosen authors. We show  a
 better fit  is
$h \sim$ $\sqrt N_c /log N_c \sim \sqrt N_p /log N_p$ (see relations (2a,b)).
It may be noted that  analysis of larger data set (see e.g., \cite{radicchi2013analysis}) indicated significantly lower value ($\simeq 0.42$) of $\gamma$. As may be seen from our analysis (see Figs. \ref{fig6} and \ref{fig7}), this is probably due to the choices of data corresponding to very large deviations from the
respective critical points to fit power law scaling form and the missing $log N$ term (in relations (2)) in the scaling 
form. It suggests, in absence of this
information, extending the data sets for
scientists who may not be that
competitive may lead to wrong conclusions.
In fact, while all the data for $h$-index
of FBM avalanches and percolation clusters
fit so well with the Widom-Stauffer
scaling (2) with $\gamma = 1/2$, when the
data for their critical points  are fitted
to $h \sim N^{\gamma}$ (see insets of Figs,
\ref{fig3} and \ref{fig5}), one gets best fits for $\gamma$
around  $0.40$ and $0.48$ respectively.

We also show that when the number $N_m$
of the members of national assemblies or
of parliaments for different countries of
the world are identified effectively as
their Hirsch index $h$, then $N_m$ indeed
would scale (see Fig. \ref{fig8}  in section IID)
with the total population $N$ as $\sqrt N
/log N$ for the respective countries.
This observation should help comprehending
the discrepancies with the proposed $N_m
\sim \sqrt N$ relationship, reported in a
very recent analysis \citep{margaritondo2021size}.

For an additional check for the $logN$ term in the scaling
behaviour of the Hirsch index, we fitted the data for $h$ in
Fiber bundle models (section IIA, for stress values  $\sigma$
near the failure point   $\sigma_c)$, for percolation systems
(section IIB, for site occupation concentrations  $p$  both
above and below the percolation threshold $p_c$), paper
citation data  (section IIC, for authors with Kolkata index
values $k$ both above and below the  critical value $k_c$ )
and parliament membership data (section IID) to the scaling
forms $h = a {\sqrt N}$  and $ h = \bar{a}\sqrt{N}/log(N)$,
where $N$ denotes the appropriate system size (total number
$N$ of fibers in the FBM, total number $N$ of lattice sites,
total number of citations $N_c$ of the author, or the total
population $N$ of the  country). The estimated best fit values of the pre-factors $a$ or $\bar{a}$ are given in Table I. The range of error bars in these estimates of $a$ and  $\bar{a}$ are such that, more than $95\%$ of the data points fall within the indicated ranges.  If we assume the value of the pre-factor to be the same
across the systems considered (Young’s asymptotic result
\cite{yong2014critique} suggested the  value of the  pre-factor  $a$ for citation
to be about $0.54$), the observed fitting error ranges for other
systems  (see Table I) then clearly suggest the fit with
$log N$  term to be more appropriate.  In fact, the average
value of the pre-factor ($\bar a \simeq 3.0$ in Table I)  fits
reasonably across the systems considered and also agrees
with the  values  of   $f(0)$ in  Figs. \ref{fig2} (for FBMs), \ref{fig4} (for
percolating systems) and \ref{fig6new} (for paper citations by an individual scientist). The comparative  agreements of the values of
$a$  (with large percentile errors) and $\bar{a}$ ($=f(0)$ in
Figs. \ref{fig2}, \ref{fig4} and \ref{fig6new} having  smaller percentile errors) in table I clearly
indicates that the fitting of $h$-index to the scaling relation (1) is much better. These errors due to the fittings
with scaling relations (2) therefore become much more
prominent for   parliament member number  data,
where the critical intervals or the country's distance from the respective critical point are completely unknown.

\begin{table}[H]
\caption{Values of the pre-factors $a$ and  $\bar{a}$ for $h-$index scaling fits.}
\begin{center}
\begin{tabular}{|c|c|c|c|c|} 
 \hline
 & FBM avalanche sizes & percolation cluster sizes & citation sizes  & parliament sizes \\
 \hline
fit to $h=a\sqrt{N}$ & & & & \\
$a=$ & $0.15\pm 0.10$ & $0.15\pm 0.10$ & $0.45\pm 0.15$ & $0.10\pm0.08$ \\
 \hline
 fit to $h=\bar{a}\sqrt{N}/log(N)$ & & & & \\
 $\bar{a}=$ & $2.00\pm 1.00$ & $2.50\pm 1.00$ & $3.50\pm 0.50$ &  $2.00\pm 1.25$ \\
 \hline
\end{tabular}
\end{center}
\end{table}

In conclusion, we have explored the best fit of the
Hirsch index values $h$ for system sizes $N$ (or $N_c$) with a
Widom-Stauffer like finite size scaling form Eq. (1). This is essentially based
on the Monte Carlo simulation data collapse in the fiber bundle model and percolating systems (for $\sigma$ or $p$ away from $\sigma_c$ or $p_c$ respectively; see Figs. \ref{fig2} and \ref{fig4}) and data analysis for citations of 100 scientists ($k$ away from $k_c$; see Fig. \ref{fig6new}) analyzed in ref. \cite{ghosh2021limiting}.  Data  for parliament member numbers (identified as the corresponding country's $h$ index; see Fig 9) are fitted to the relation Eq. (2), as the equivalent critical interval ($[\sigma_c-\sigma]$, $|p_c-p|$ or $|k_c-k|$) are unknown. We find the scaling fit  (to relations Eq. (1) and (2)) deteriorate considerably if the $log N$ term is dropped. We give  in the table I the estimated error in the pre-factors $a$ and $\bar{a}$, and assuming the pre-factor to have the same value across the systems considered here, we again find the finite size scaling relation for the Hirsch index $h$  with $log N$ term ($h=\bar{a} \sqrt{N}/log N$, $\bar{a}\equiv f(0)$ of the Widom-Stauffer relation Eq. (1) for finite system size $N$) to be more appropriate for the scaling behaviour  of Hirsch index near the respective critical points.

\section*{acknowledgments}
We are grateful to Parongama Sen  for useful comments and suggestions. BKC is thankful to the Indian National Science Academy for their Senior Scientist Research Grant.


\end{document}